\documentclass[letterpaper]{jpconf}
\usepackage{graphicx}
\usepackage[square,sort&compress]{natbib}
\begin{document}
\title{Crystal field studies on the heavy fermion compound CeNi$_8$CuGe$_4$}

\author{L Peyker$^1$, C Gold$^1$, E-W Scheidt$^1$, H Michor$^2$, T Unruh$^3$, A Senyshyn$^{3,4}$, O Stockert$^5$ and W Scherer$^1$}

\address{$^1$ CPM, Institut f\"{u}r Physik, Universit\"{a}t Augsburg, 86159
Augsburg, Germany}
\address{$^2$ Institut f\"{u}r Festk\"{o}rperphysik, Technische Universit\"{a}t Wien,
1040 Wien, Austria}
\address{$^3$ Forschungsneutronenquelle Heinz Maier-Leibnitz, Technische Universit\"{a}t M\"{u}nchen,
85747 Garching, Germany}
\address{$^4$ Strukturforschung, Materialwissenschaft, Technische Universit\"{a}t Darmstadt,
64287 Darmstadt, Germany}
\address{$^5$ Max-Planck-Institut CPfS, 01187 Dresden, Germany}

\ead{ludwig.peyker@physik.uni-augsburg.de}

\begin{abstract}
Substitution of nickel by copper in the heavy fermion system
CeNi$_{9-x}$Cu$_x$Ge$_4$ alters the local crystal field environment
of the Ce$^{3+}$-ions. This leads to a quantum phase transition near
$x\approx0.4$, which is not only driven by the competition between
Kondo effect and RKKY interaction, but also by a reduction of an
effectively fourfold to a twofold
degenerate crystal field ground state.
To study the consequences of a changing crystal field in
CeNi$_8$CuGe$_4$ on its Kondo properties, inelastic neutron
scattering (INS) experiments were performed. Two well-defined
crystal field transitions were observed in the energy-loss spectra
at 4\,K. The crystal field level scheme determined by neutron
spectroscopy is compared with results from specific heat
measurements.
\end{abstract}

\section{Introduction}
The heavy fermion system CeNi$_9$Ge$_4$ exhibits an outstanding
fermi-liquid behavior with the largest value of the electronic
specific heat $\Delta C/T\approx5.5$\,Jmol$^{-1}$K$^{-2}$ ever
recorded in the absence of any magnetic order \cite{Michor2004,
Killer2004}. The crystal field, investigated by single crystal
susceptibility analysis, consists of a quasi-quartet ground state
and one doublet at $\Delta _2\simeq11$\,meV
($\mathrel{\widehat{=}}128$\,K) \cite{Michor2006}. The quasi-quartet
is internally split into two doublets separated by $\Delta
_1\simeq0.5$\,meV ($\mathrel{\widehat{=}}6$\,K). The
Kondo-temperature $T_{\mathrm K}\simeq3.5$\,K, determined by
inelastic neutron scattering studies, falls in the same
energy-range as the splitting of the quasi-quartet ground state.\\
While this unusual fermi-liquid behavior is mainly driven by
single-ion effects \cite{Killer2004}, Ni/Cu substitution leads to
antiferromagnetic order in CeNi$_8$CuGe$_4$ with $T_{\mathrm
N}=175$\,mK. Profound analysis of specific heat, magnetic
susceptibility and electrical resistivity of the substitution-series
CeNi$_{9-x}$Cu$_x$Ge$_4$ reveals a continuous tuning of the ground
state from an effectively fourfold degenerate nonmagnetic Kondo
ground state in CeNi$_9$Ge$_4$ towards a magnetically ordered,
effectively twofold degenerate ground state in CeNi$_8$CuGe$_4$
\cite{Peyker2009}. Around the critical Cu-concentration of $x \simeq 0.4$
the system exhibits quantum critical behavior with $\chi$ and
$C/T\propto\ln T$ and $\rho\propto T$. Previous investigations on
the system suggested, that this quantum phase transition is not only
driven by the competition between Kondo-effect and RKKY interaction,
but also by a reduction of the effective crystal field ground state
degeneracy, caused by two or more different crystal field
environments of the Ce$^{3+}$-ion \cite{Peyker2009}. To prove these
suggestions, which were mainly based on specific heat measurements,
we performed neutron scattering experiments on the compound
CeNi$_8$CuGe$_4$ in order to study the crystal field and the
disorder effects, occurring in the chemical environment of the
Ce$^{3+}$-ion.

\section{Experimental Details}
Several polycrystalline samples of CeNi$_8$CuGe$_4$ and of the
nonparamagnetic reference compound LaNi$_8$CuGe$_4$ have been
synthesized by arc melting of pure elements in a highly purified
argon atmosphere. The samples were flipped over several times and
remelted to obtain the highest possible homogeneity. After annealing
the samples under vacuum for one week at 950$^{\circ}$\,C, X-ray
diffraction was used to check the phase purity.\\
Neutron scattering studies were performed at the research reactor
FRM-II (Garching, Germany), where inelastic scattering studies were
carried out on the time-of-flight spectrometer TOFTOF, whilst
elastic scattering was studied at the high-resolution structure
powder diffractometer SPODI ($\lambda=1.5483 $\,{\AA}).
For INS studies, about 35\,g of each compound were filled in a thin
double-walled aluminium can mounted onto a closed-cycle
refrigerator, whereas for powder diffraction experiment on
CeNi$_8$CuGe$_4$ a single-wall cylindric vanadium container (14 mm
in diameter, 0.15 mm
wall thickness) was used.\\
Powder X-ray diffraction experiment showed that both
\begin{figure}
\begin{center}
\includegraphics[width=24pc]{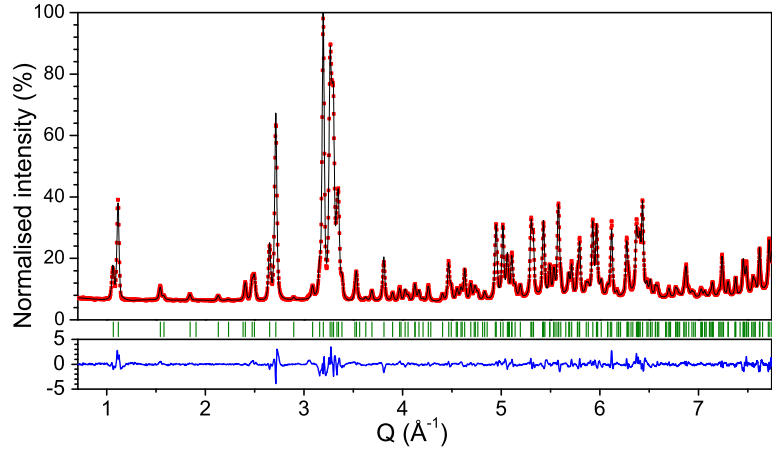}
\end{center}
\caption{\label{fig1}(Color online) Results of Rietveld refinement
for neutron powder diffraction data at 4\,K for CeNi$_8$CuGe$_4$.
Experimental data are shown by points, the continuous line denotes
the calculated profile and the lower plot represents the difference
between the observed and calculated intensity. Calculated positions
of Bragg reflections are shown by vertical tick marks.}
\end{figure}
CeNi$_8$CuGe$_4$ and LaNi$_8$CuGe$_4$ are single-phase materials and
their structure belongs to the LaFe$_9$Si$_4$-type (space group
$I4/mcm$) with three Ni-sites (Ni1:$16k$, Ni2:$16l$, Ni3:$4d$), one
Ge ($16l$) and one Ce site ($4a$). It is evident that
X-ray diffraction did not revealed any ordering of copper in the
lattice due to the minute differences in the scattering factors of
Ni and Cu. In order to determine which of the individual Ni-sites is
preferentially substituted by Cu and to estimate the degree of Ni/Cu
disorder in CeNi$_8$CuGe$_4$ a neutron diffraction experiment was
performed at 300\,K and 4\,K. First, the sample homogeneity was
checked by a full profile decomposition technique, where the
background (determined by a linear interpolation between points in
non-overlapping regions), lattice, profile (Pseudo-Voigt function)
and asymmetry parameters were refined. A rather weak, but still
noticeable reflection broadening was observed over the entire
$Q$-range accessible in the diffraction experiment. After
deconvolution with the instrumental resolution function this line
broadening could be assigned to a size-effect, which was best
simulated with platelet-like shaped crystallites (ca. 400 {\AA} in
diameter and surface parallel to [200]). In further structure
considerations, based on the Rietveld method, observed weak
microstructural effects were included by simultaneous refinements of
the Caglioti profile parameters. For the localization of the copper
atoms in the lattice, the occupancies of the different Ni sites were
checked. As a result of the best Rietveld fit (Fig.\,\ref{fig1})
99\% of the copper atoms could be located on the Ni1 site. This is
in good agreement with bandstructure calculations based on the new
full-potential augmented spherical wave method \cite{Peyker2009},
which clearly show an energetic preference for a Ni/Cu substitution
at the Ni1 site. Furthermore, a little excess of the occupancy on
the $16l$-place of both Ni and Ge was found, which is probably related to a
weak static disorder existing in the material.\\
For the determination of the crystal field level scheme an incident
neutron energy of $E_{\mathrm i}=16.9$\,meV has been chosen at a
sample-temperature of 4\,K. Additionally, the nonparamagnetic
reference compound LaNi$_8$CuGe$_4$ and a vanadium standard have
been measured to determine the phonon
\begin{figure}
\begin{minipage}{18pc}
\begin{center}
\includegraphics[width=15pc]{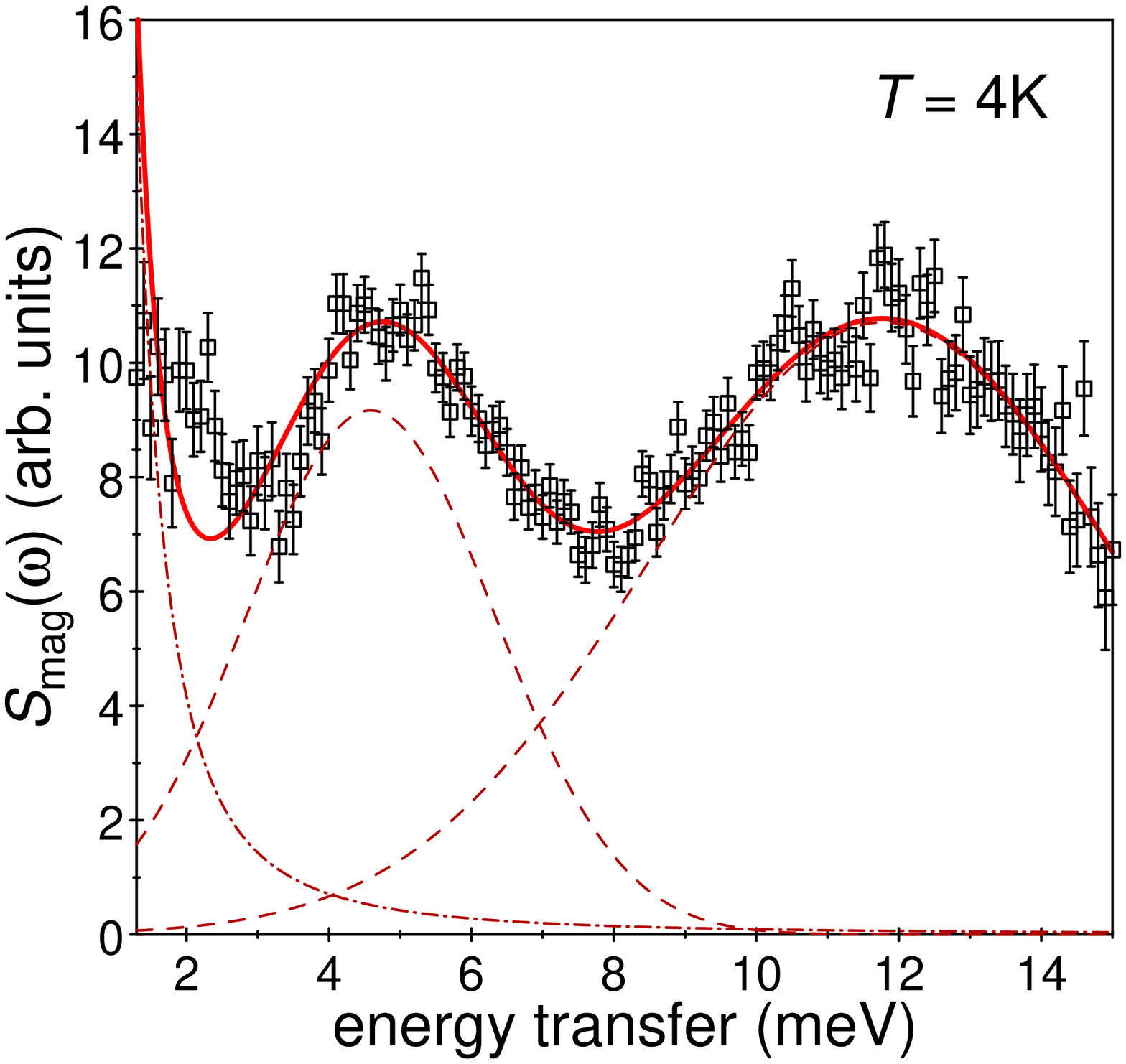}
\end{center}
\caption{\label{fig2}(Color online) Magnetic scattering function
$S_{\mathrm{mag}}(Q,\omega)$ of CeNi$_8$CuGe$_4$ at 4\,K. The solid
line represents a fit to a spectral function consisting of two
Gaussians (dashed line) and one Lorentzian (dash-dotted line).}
\end{minipage}\hspace{2pc}%
\begin{minipage}{18pc}
\begin{center}
\includegraphics[width=15pc]{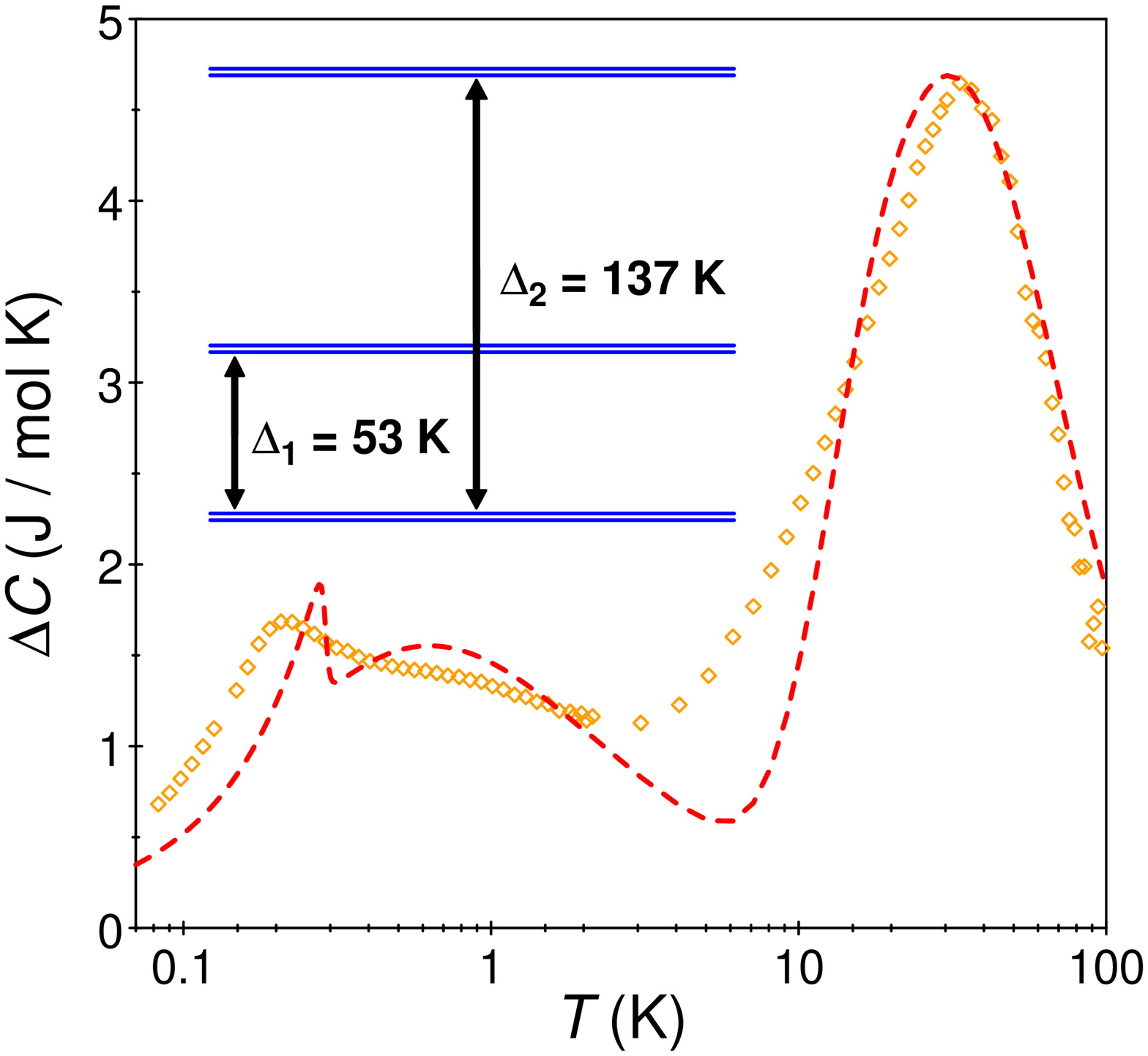}
\end{center}
\caption{\label{fig3}(Color online) The magnetic contribution of the
specific heat $\Delta C$. The dashed line is a theoretical
adjustment to the data taking into account the determined crystal
level scheme displayed in the insert.}
\end{minipage}
\end{figure}
part of the scattering and the detector efficiency.
The main challenge in the quantitative interpretation of measured
INS spectra is to derive the phonon contribution correctly,
particularly in compounds, such as CeNi$_8$CuGe$_4$, that have a
large nuclear cross section. The phonon correction used in our case,
was based on the different $Q$-dependencies $(Q=\frac{4\pi}{\lambda}
\sin\Theta)$ of the phononic
(approximately $\propto Q^2$) und magnetic scattering (form-factor
decrease with increasing $Q$-values). Therefore, we summed the
detector counts at low
($\langle\Theta_\mathrm{low}\rangle=20^\circ$) and high
($\langle\Theta_\mathrm{high}\rangle=125^\circ$) scattering angles
with the corresponding momentum transfer ranges of
$Q_\mathrm{low}=0.4-1.6$\,{\AA}$^{-1}$ and
$Q_\mathrm{high}=4.8-5.4$\,{\AA}$^{-1}$. The nonparamagnetic reference compound was used to
determine the ratio of the high-angle and low-angle scattering
intensities. This ratio is used to scale the high $Q$ to the low $Q$
response in CeNi$_8$CuGe$_4$ and to subtract the phonon contribution
\cite{Murani1994}. In Fig.\,\ref{fig2} the corrected data, which
show the magnetic scattering function $S_{\mathrm{mag}}(Q,\omega)$
of CeNi$_8$CuGe$_4$, are displayed. Two broad excitations around
4.5\,meV and 12\,meV are observed in the spectrum, which can be
assigned to magnetic transitions. The deviation from the fit around
2\,meV could be caused by disorder effects as already mentioned in
\cite{Peyker2009}. In order to analyze the scattering function in a
proper way, we fitted a combination of two Gaussian-functions
(dashed line), representing the crystal field transitions, and a
Lorentzian-function (dash-dotted line), describing the elastic
scattering line, to our data. From the fit the values of the crystal
field splitting $\Delta _1\approx4.6\pm0.2$\,meV
($\mathrel{\widehat{\approx}} 53$\,K) and $\Delta
_2\approx11.8\pm0.2$\,meV ($\mathrel{\widehat{\approx}} 137$\,K) can
be determined.

\section{Discussion and conclusions}

The crystal field scheme determined from the INS experiments
described above is displayed in the insert in Fig.\,\ref{fig3}. In
contrast to the preliminary results published in \cite{Peyker2009},
where the phonon-contribution was taken into account by a simple
La-subtraction and which conclude that more than one Ce site with
different environments exist, we observe only one Ce site with
$\Delta_1\approx53$\,K and $\Delta_2\approx137$\,K. In
Fig.\,\ref{fig3} an analysis of the magnetic contribution to the
specific heat $\Delta C$ of CeNi$_8$CuGe$_4$ considering the above
determined crystal field scheme is shown. The interplay between
molecular field and Kondo effect within the doublet ground state is
regarded by the resonant-level model in combination with a molecular
field approach (for details see \cite{Peyker2009}). The good
agreement with the experimental data support the investigated
crystal field scheme and therefore the fact, that the system
CeNi$_8$CuGe$_4$ exhibits only one crystal field environment of the
Ce-atoms with the Cu-atoms localized at the Ni1 site. The deviation
between the experimental data and the theoretical adjustment around
7\,K could be due to a weak static disorder, which can not be
excluded, especially because of the observation of the additional
scattering intensity in the INS spectra around 2\,meV. Structural
disorder can indirectly be confirmed by relatively high magnitudes
of isotropic displacement parameters at 4 K as well as a weak excess
occupation on Ni2 $(16l)$ and Ni1 $(16k)$ atomic sites.\\
Comparing the crystal field scheme with the pure Ni-compound
CeNi$_9$Ge$_4$, where the crystal field consists of a quasi-quartet
ground state at around $\Delta_1\approx6$\,K and one doublet at
$\Delta_2\approx128$\,K, the substitution of Ni by Cu leads to a
splitting of the quasi-quartet into a twofold degenerated ground
state and a higher shifting doublet, while the excited doublet
remains at almost the same energy. This reduction of the effective
crystal field ground state degeneracy could be related to the
quantum phase transition in CeNi$_{9-x}$Cu$_x$Ge$_4$. So not only to
the competition between Kondo effect and RKKY interaction but also
the change in the crystal field could be the driving force in this
quantum critical scenario. To prove that this crystal field shift is
a crucial tuning parameter in terms of the observed quantum critical
transition, further INS experiments on other Ni/Cu-composition are
in progress, to study the change of the crystal field in detail,
especially around the quantum critical transition ($x=0.4)$ itself.

\section{Acknowledgments}

This work was supported by the Deutsche Forschungsgemeinschaft (DFG)
under Contract No. SCHE487/7-1 and by the COST P16 ECOM project of
the European Union.


\begin{thebibliography}{1}
\expandafter\ifx\csname url\endcsname\relax
  \def\url#1{{\tt #1}}\fi
\expandafter\ifx\csname urlprefix\endcsname\relax\def\urlprefix{URL }\fi
\providecommand{\eprint}[2][]{\url{#2}}

\bibitem{Michor2004}
Michor H, Bauer E, Dusek C, Hilscher G, Rogl P, Chevalier B, Etourneau J,
  Giester G, Killer U and Scheidt E-W {2004} {\em {J. Magn. Magn. Mater.}\/}
  {\bf {272}} {227--228} ISSN {0304-8853}

\bibitem{Killer2004}
Killer U, Scheidt E-W, Eickerling G, Michor H, Sereni J, Pruschke T and Kehrein
  S {2004} {\em {Phys. Rev. Lett.}\/} {\bf {93}} ISSN {0031-9007}

\bibitem{Michor2006}
Michor H, Adroja D~T, Bauer E, Bewley R, Dobozanov D, Hillier A~D, Hilscher G,
  Killer U, Koza M, Manalo S, Manuel P, Reissner M, Rogl P, Rotter M and
  Scheidt E-W {2006} {\em {Physica}\/} {B} {\bf {378-80}} {640--643} ISSN
  {0921-4526}

\bibitem{Peyker2009}
Peyker L, Gold C, Scheidt E-W, Scherer W, Donath J~G, Gegenwart P, Mayr F,
  Eyert V, Bauer E and Michor H {2009} {\em {J. Phys.: Condens. Matter}\/} {\bf
  {21}} {235604}

\bibitem{Murani1994}
Murani A~P {1994} {\em {Phys. Rev.}\/} {B} {\bf {50}} {9882--9893} ISSN
  {0163-1829}

\end{thebibliography}
\providecommand{\newblock}{}

\end{document}